\documentclass[dvips,10pt]{article}
\usepackage{moriond,epsfig}

\def\be{\begin{equation}}
\def\ee{\end{equation}}
\def\bea{\begin{eqnarray}}
\def\eea{\end{eqnarray}}

\begin{document}
\vspace*{4cm}
\title{EXPERIMENTAL PARTICLE DARK MATTER SEARCH}

\author{M. LOIDL}

\address{CEA Saclay, DSM/DAPNIA/SPP, 91191 Gif-sur-Yvette Cedex, France}

\maketitle\abstracts{
After introducing several aspects of the motivation for particle dark matter search,
experimental principles and the present state of the main experiments are presented.
Direct searches for WIMPs are explained in some detail; indirect WIMP searches
and axion searches are presented more briefly.}

\section{Introduction}

All viable models in present-day cosmology require the existence of non-baryonic 
cold dark matter with a cosmological density dominating over all other forms of matter 
in the universe.

The generally accepted inflationary cosmological models require the universe to be 
flat, i.~e. the total matter and energy density $\Omega = 1$ in units of the critical 
density separating a universe with positive from one with negative curvature of 
space-time, to avoid the ``fine-tuning problem'': any deviation of the cosmological 
density from unity would have increased by many orders of magnitude during inflation, 
contradicting observations \cite{Boerner93}. Thus $\Omega$ should be precisely equal
to one from the beginning.
The flatness of the universe is confirmed, $\Omega = 1$ to 
within a few per cent, by measurements of the first acoustic oscillations in the 
angular power spectrum of the cosmic microwave background (CMB) with the BOOMERANG, 
MAXIMA, DASI and other mostly satellite or balloon borne instruments 
\cite{Masi02,Stompor01,Pryke02}.

The observation of the brightness-to-distance relationship of type Ia supernovae 
reveals evidence for the presence of some form of Dark Energy $\Lambda$ with a density 
$\Omega_\Lambda \approx 0.7$ \cite{Perl99}. A similar value (0.65 - 0.85) has been 
obtained by a 
combined analysis of CMB data and the power spectrum of the matter density 
distribution \cite{Efs01}. This leaves an overall matter density $\Omega_m \approx 0.3$, 
in good 
agreement with the value inferred from the dynamics of galaxies within 
clusters. On the other hand, the theory of big bang 
nucleosynthesis combined with measurements of the cosmic abundances of helium and,
more recently, deuterium relative to that of hydrogen limit the cosmic density of 
baryonic matter quite 
strictly to $\Omega_b h^2 = 0.020 \pm 0.002$, where $h$ denotes the Hubble parameter
\cite{Burles01}. 
Also this value agrees well with that resulting from CMB measurements.

Thus the missing matter must be in form of non-baryonic particles which interact only 
weakly with normal matter. Most particles in the universe have frozen out of thermal 
equilibrium with the primordial plasma at a temperature corresponding to their 
respective mass. 
Dark matter candidates that were relativistic at the time of freeze-out are referred 
to as hot dark matter (HDM); those which were non-relativistic as cold dark matter 
(CDM). This dark matter ``temperature'' can be ``measured'' by means of the power 
spectrum of the matter density distribution as it results from all-sky galaxy surveys. 
HDM would have, due to its relativistic streaming while structure formation in the 
early universe, wiped out small-scale structures,
i. e. prohibited the early formation of individual galaxies. CDM instead, freezing out 
earlier 
than baryons due to its weak interaction, could have fallen into density fluctuations 
very early thus enhancing the formation of small scale structures. The observed 
structures are best described by the matter content being dominated by CDM, with some 
HDM being allowed.
A small HDM component can readily be explained by low-mass neutrinos since neutrinos 
are now, after the intriguing results from Superkamiokande and SNO 
\cite{Fukuda99,Ahmad01}, generally 
considered having a small mass: the present neutrino mass limits correspond to 
a contribution to the cosmic density in the range $0.001 < \Omega_\nu < 0.18$ 
\cite{Ahmad01}. 

This article will concentrate on CDM candidates and 
approaches to their detection, and within this frame it will be further constrained 
to those candidates that are well motivated in that they emerge from well-established 
theories in particle physics and can at the same time constitute the required matter 
density without arbitrary assumptions of their physical parameters: supersymmetric 
Weakly Interacting Massive Particles (WIMPs) and axions.

\section{SuSy WIMPs}

Supersymmetry introduces a global symmetry between the fermionic and the bosonic 
sectors of the thus extended standard model, ascribing to each fermion a supersymmetric 
bosonic partner and vice versa. In most supersymmetric models with conservation 
of the newly introduced quantum number R-parity the lightest supersymmetric particle 
(LSP) is stable. In the minimal supersymmetric extensions of the standard model
(MSSM) the LSP is the lightest neutralino $\chi$, a linear combination of the 
superpartners of the gauge and Higgs bosons: bino, wino and the two neutral 
higgsinos. The neutralino is an electrically neutral Majorana fermion with a mass range 
between few GeV and few TeV and a WIMP-nucleon cross section 
in the range $10^{-10}$\,pb\,$\leq 
\sigma_{\chi-n} \leq 10^{-5}$\,pb. Accelerator searches set a lower mass limit $m_\chi 
\geq 45$\,GeV \cite{Lak02}. 

Standard assumptions for WIMP dark matter searches are that WIMPs are gravitationally 
bound to the galaxy forming an isothermal spherical halo with a local density of 0.3 
GeV/cm$^3$ and a Maxwellian velocity distribution with a mean value of 270\,km/s, 
truncated at the escape velocity from the galaxy of 650\,km/s.

Beyond the MSSM discussed here, other SUSY models put into play several other DM particle 
candidates like s-neutrinos, axinos, self-interacting DM, wimpzillas, simpzillas, cryptons, 
Q-balls, see e.\,g. ref. \cite{Berg02} and references cited therein.

\section{Direct WIMP searches}

The exclusive possibility to directly detect WIMPs is to measure the small amount of energy 
(few to few ten keV) deposited in a detector when a WIMP scatters elastically off a 
nucleus of the detector's target material. Due to the small WIMP-nucleon cross 
section event rates are expected in a range below one event per day and per kg of 
detector mass. Therefore, evidently, a large mass is required to have a realistic 
chance to observe WIMP signals, in conjunction with a very low detection threshold. 

Both scalar, i.e. spin-independent, and axial, i.e. spin-dependent, coupling of WIMPs 
to nucleons must be considered. If the coupling is predominantly scalar, the 
coupling to nuclei is coherent, and the WIMP-nucleus cross section roughly 
proportional to $A^2$. In this case heavy target nuclei strongly enhance the detection 
efficiency. In the case of axial 
coupling the WIMP couples to the unpaired spin in a nucleus with non-zero,
half-integral net spin. 
This requires a high natural abundance or enrichment of isotopes with 
half-integral spin in the target material.

The main challenge for WIMP searches arises from the low count rate, demanding a very
low radioactive and cosmic ray background. Careful selection of radiopure materials 
for the experimental set-up and very efficient shielding against radioactivity from
the laboratory environment are needed as well as the installation of the experiment 
in a deep underground site. Remaining radioactive background causes mostly electron
recoils, hence the detector should allow to discriminate electron recoils from the
nuclear recoils caused by WIMPS.
Neutrons however  scatter 
off nuclei and can't be distinguished from potential WIMP events (if they scatter only 
once - multiple scatter events in a segmented detector can obviously be 
attributed to neutrons). This implies for 
experiments of ultimate sensitivity an efficient neutron shield as well as a muon 
veto since the few muons from secondary cosmic radiation that penetrate even into the 
deepest underground sites can induce neutrons inside the neutron shield. A typical set-up 
uses several tons (10 - 20\,cm) of lead shield, with an inner layer of archeological 
lead or high purity copper, 30 - 50\,cm of paraffine or polyethylene, possibly a 
cadmium sheet, a nitrogen-flushed box or plastic bag to keep off radon, and a plastic 
scintillator muon veto.

A relatively simple rejection of electron recoils by pulse shape discrimination (PSD) 
relies on 
different signal decay times in scintillators for electron and nuclear recoils, 
respectively, reaching low to moderate rejection capabilities. 
In the low energy range where WIMP signals are expected the difference in decay time
becomes very small. More efficient 
rejection can be realised using two detection channels for two types of 
excitations created by a particle interaction in the detector target: Scintillation 
and ionization, scintillation and phonons or ionization and phonons. Typically two 
types of excitations have different quenching factors: a quenching factor expresses 
a lower detectable energy yield for nuclear recoils than for electron recoils of the 
same deposited energy. Hence, the ratio of signal amplitudes in the two detection 
channels provides a powerful means to distinguish the two interaction types.

Positive evidence for a WIMP signal could arise from the kinematics of the earth 
within the presumably non-rotating WIMP halo. The sun is orbiting about the galactic 
center with a velocity of $\sim 220$\,km/s. The resulting relative sun-halo velocity is 
modulated by the earth's orbit about the sun at $\sim 30$\,km/s. An annual modulation 
of the count rate by few per cent could yield positive signature. The most convincing 
evidence would arise from a modulation of the directionality of the 
recoil nuclei corresponding to the above mentioned kinematics, given a direction 
sensitive detector. In this case the directionality is additionally modulated 
diurnally by the earth's rotation about its axis.

\subsection{Classical germanium detectors}

For several years the Heidelberg-Moscow Experiment, using classical Ge detectors at 
liquid nitrogen temperature (77\,K) and originally designed to search for neutrinoless 
double-beta decay, achieved the lowest background (0.05\,ev/kg/keV/day above 10\,keV) 
and the highest sensitivity of any dark 
matter search \cite{Baudis98}. This was possible due to the available high purity of 
Ge and a 
substantial, fundamental work on low background conditions. The Heidelberg group is 
presently operating in the Gran Sasso underground lab a 200\,g p-type Ge detector 
surrounded by a 2\,kg n-type Ge veto detector in a well arrangement (HDMS) \cite{Baudis01}. 
Photons or 
neutrons interacting in the inner detector have a considerable probability to 
generate a coincident signal in the veto detector. Background level and sensitivity 
are comparable to the previous experiment. In the GENIUS project \cite{Klapdor01} 
proposed by the same 
group 100\,kg of unencapsulated Ge crystals shall be surrounded by a large 
volume of highest purity liquid nitrogen to assure a very radiopure surrounding as 
well as shielding from external radioactivity. A test facility for this project has 
recently been approved.

The IGEX detector, deployed in the Canfranc tunnel in the Pyrenees, 
has reached the highest dark matter sensitivity ($\sigma_{\chi-n} \approx 7 
\times 10^{-6}$\,pb) of all experiments using classical Ge detectors, excluding the 
upper part of the DAMA region (see section~\ref{subsec:scint}) \cite{Morales02}. It 
consists of up to three 
2\,kg Ge crystals enriched to 86\,\% in $^{76}$Ge and has a lower energy threshold than 
the experiments described above.

The principal problem of these experiments is the lack of any background 
discrimination capability. Even at the lowest background level some events will occur 
which will never be identifiable as electron or nuclear recoils. 

\subsection{Scintillator detectors}\label{subsec:scint}

Scintillators constitute the second class of classical detectors in use for dark 
matter search. Scintillation in both solids and liquids is a well established 
detection technique and large detector masses are readily installed. 

\begin{figure}
\begin{center}
\epsfig{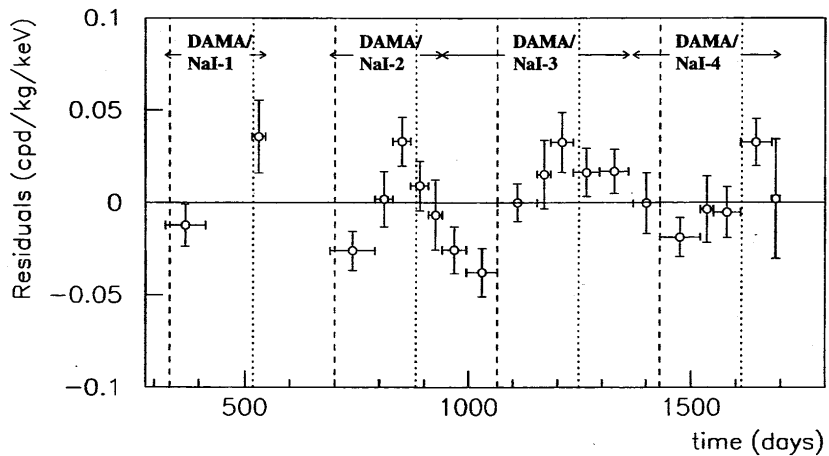}
\caption{Annual modulation of the count rate as observed by the DAMA collaboration
over a period of four years.}
\label{fig:DAMA}
\end{center}
\end{figure}

The DAMA collaboration is operating $\sim$ 100\,kg of NaI crystals, each viewed by two 
photomultiplier tubes (PMTs), in the Gran Sasso underground lab and has collected 
within four years $\sim$ 58000\,kg$\cdot$days of statistics. The non-zero spin of the sodium 
nucleus makes these detectors also sensitive to axial coupling. The DAMA group  
claims to observe annual modulation of their count rate, as shown in fig.~\ref{fig:DAMA}, 
which is compatible in 
amplitude and phase with a signal from WIMPs of (52$^{+10}_{-8}$)\,GeV mass and a 
WIMP-nucleon cross section of $(7.2^{+0.4}_{-0.9}) \times 10^{-6}$\,pb \cite{Bernabei00}, 
see also fig.~\ref{fig:EDW}(b). Background 
rejection is based on PSD, using a statistical method which is less efficient than an
event-by-event discrimination. Moreover, 
scintillators suffer from a relatively large quenching factor, hence the threshold 
for nuclear recoils is several times higher than that of ``visible'' energy. The 
major part of the data interpreted as a WIMP signal lies in the first (2 - 3\,keV) 
energy bin above the PMT noise where PSD seems questionable. Despite these difficulties the 
observed annual modulation amplitude corresponds to the most optimistic value that 
could be expected for a pure WIMP signal. Severe criticism has arisen in the 
community \cite{Gerbier99,Spooner98} ascribing the observed annual modulation rather 
to systematics than to a WIMP signature.

Background rejection by PSD is remarkably more efficient in liquid xenon scintillator 
detectors. This allows the UKDMC group to reach competitive dark matter sensitivity,
excluding the upper half of the DAMA region \cite{Hart02},  
already with low statistics collected within a few months with a 4\,kg prototype 
detector, ZEPLIN-I, mounted in the Boulby mine in the UK. The scintillator is viewed by three 
PMTs which allow, by triple coincidence, to reject spurious events in or near the 
PMTs. A liquid scintillator Compton veto and a lead shield complete the set-up. The ZEPLIN 
group is presently developing two even more efficient background rejection techniques 
that allow discrimination on an event-by-event basis. They are both exploiting 
ionization occurring in the liquid Xe in addition to scintillation. In the simpler detector 
design the ionization charges are drifted within the liquid and generate a 
delayed secondary scintillation signal. In a more 
sophisticated two-phase liquid and gas detector ionization electrons are extracted 
from the liquid into the gas phase, where they are accelerated and generate 
electroluminescence that is registered as a delayed signal by the same two PMTs that 
are viewing the liquid Xe. Since nuclear recoils generate much less ionization than 
electron recoils the secondary signal is is much smaller for nuclear than for 
electron recoils. A one ton project is hoped to cover a large 
fraction of the WIMP parameter space \cite{Spooner02}.

The ELEGANT V and VI experiments based in the Japanese
Oto Cosmo Observatory are employing, beside a 730\,kg NaI crystal 
array, 7.2\,kg of CaF$_2$ scintillators \cite{Yoshida00,Hazama01}. 
The $^{19}$F nucleus offers a non-zero spin and a very
advantageous form factor which makes it particularly sensitive for axial 
coupling. Unfortunately both materials in these experiments are not of the highest 
purity.

\subsection{Cryogenic detectors}

A third class of detectors, operated at temperatures of $\sim$ 10 - 100\,mK, is using 
phonons as principal excitations and detecting them as temperature pulses. Phonons 
offer two advantages as compared to ionization and scintillation: firstly, almost the 
entire energy of nuclear recoils is transformed into phonons, whereas ionization and 
scintillating detectors have relatively high quenching factors. Secondly, the 
excitation energy of phonons, typically meV, is roughly three orders of magnitude 
lower than that of electron-hole pairs or scintillation photons. These two advantages 
result in a considerably lower energy threshold as well as in a higher energy 
resolution. Combining phonon detection with either ionization or scintillation 
measurement allows very efficient event-by-event background rejection.

The very low temperatures are required since the phonons are detected as thermal 
signals: thermal noise is strongly suppressed at such low temperatures, 
and heat capacities go dramatically down with temperature: linearly for 
metals, and even as $T^3$ for dielectrics. 

\begin{figure}[b!]
\begin{center}
\begin{tabular}{ccc}
\mbox{\epsfig{file=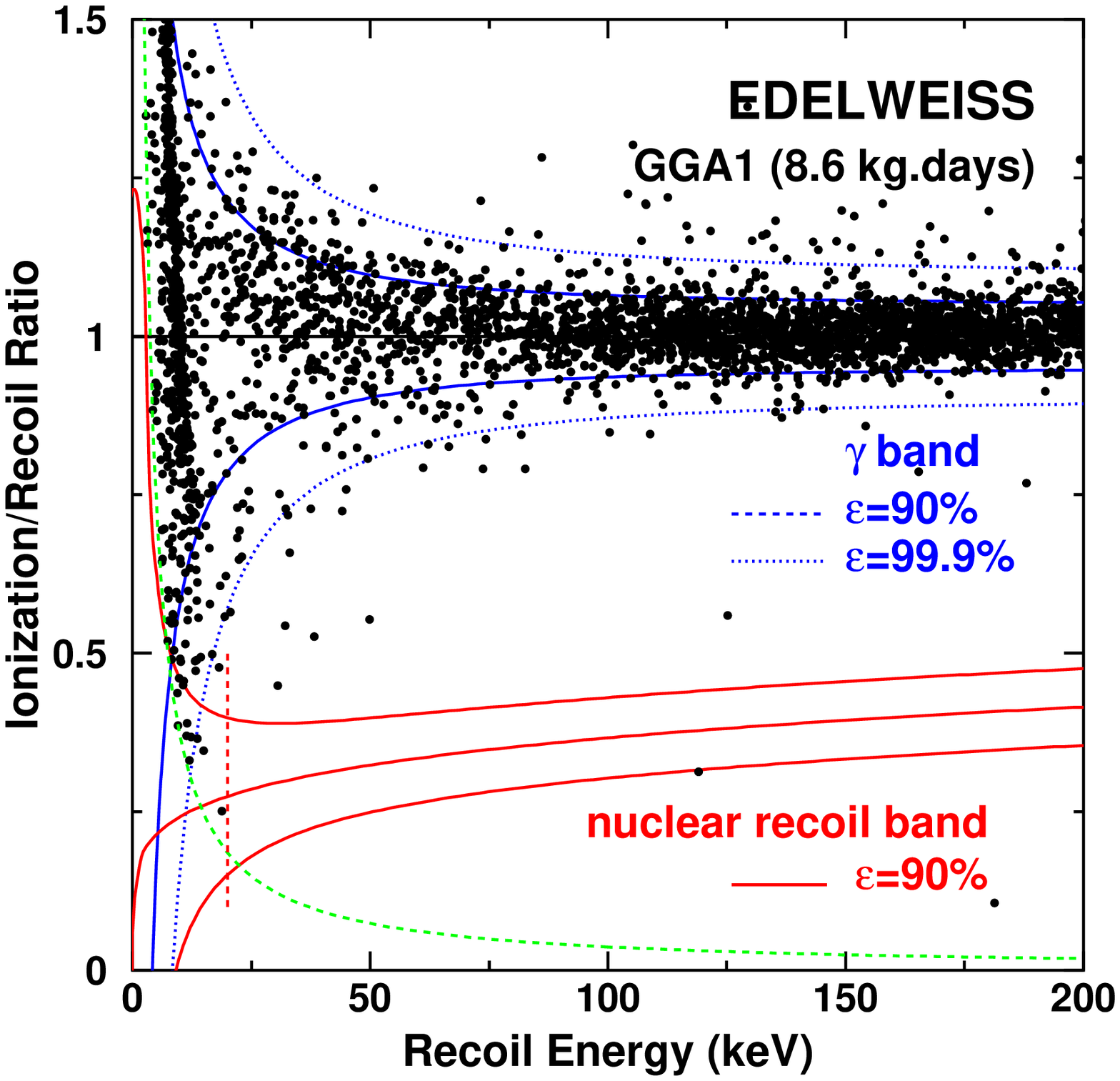,width=7.4cm}} &
\mbox{\hspace{0.5cm}} &
\mbox{\epsfig{file=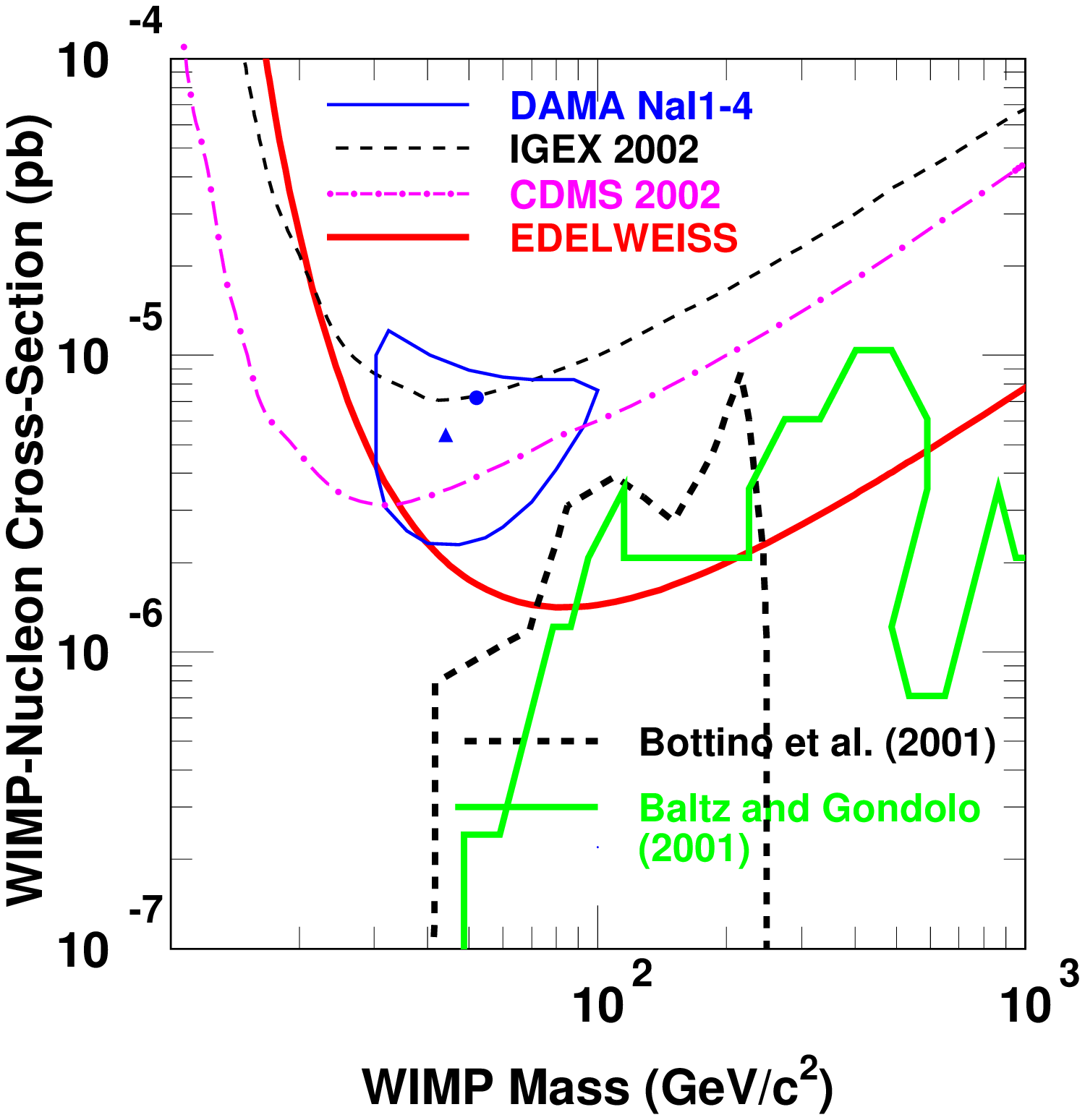,width=7.0cm}}
\end{tabular}
\caption{(a) Low background data of the EDELWEISS experiment. Almost all events are 
clearly electron recoils. WIMP and neutron events would appear in the lower nuclear 
recoil band that has been determined in calibration experiments with a neutron source.
Due to the absence of events in the 90\,\% acceptance nuclear recoil band no background
subtraction was required to derive a dark matter exclusion limit.
(b) Exclusion limits of some of the 
presently most sensitive direct WIMP searches. The region above each curve in the 
parameter space WIMP-nucleon cross section vs. WIMP mass has been
excluded by the respective experiments. The closed contour corresponds to the 
3\,$\sigma$ allowed region of the DAMA experiment, its central value is represented
by the full circle. The regions in the lower half delimited by the bold lines represent
the parameter space predicted by different SuSy model calculations.}
\label{fig:EDW}
\end{center}
\end{figure}

The CRESST experiment \cite{CRESST02} in the Gran Sasso underground lab has used in its
recently finished first phase 262\,g sapphire (Al$_2$O$_3$) crystals and superconducting 
thin tungsten films on 
the crystal surface as temperature sensors. The detectors are operated at a temperature 
within the superconducting phase transition, $\sim$\,15\,mK, where the film's electrical 
resistance is an extremely sensitive measure for temperature variations. The low resistances 
($\sim 0.1\,\Omega$) of the films are read out with SQUIDs. These detectors have 
extremely low energy thresholds (580\,eV) and very good energy resolution 
(130 - 330\,eV for 1.5\,keV X-rays) and reached the highest sensitivity for both scalar 
and axially coupled WIMPs with masses below 5\,GeV. With their low-mass Al and O 
nuclei and without background rejection, these detectors were not competitive in the 
higher WIMP mass range favoured by theory and accelerator mass limits. 

For a second phase the CRESST group has developed detectors with scintillating 300\,g 
CaWO$_4$ target crystals and superconducting W thermometers for phonon detection.
A separate small cryogenic detector measures 
scintillation light, the high quenching factor allowing 
an impressive rejection power. Up to 10\,kg of CaWO$_4$ 
detectors will be mounted in 
the existing cryogenic set-up in phase II of the experiment \cite{CRESST01}.

The CDMS collaboration operates Ge and Si cryogenic detectors with both 
superconducting thin films as described above and more conventional NTD Ge 
thermistors as temperature sensors \cite{CDMS}. Additionally, ionization charges are 
drifted to 
thin film electrodes on the crystal surfaces and provide efficient electron recoil 
discrimination. With one 100\,g Si and four 165\,g Ge detectors run in 1998 and 1999 
CDMS could cover a large part of the 3$\sigma$ allowed region of the DAMA evidence
(see fig.~\ref{fig:EDW}(b)), 
excluding the DAMA most likely candidate at $> 99.9$\,\% CL \cite{CDMS}. The experiment is 
however situated in 
a shallow underground site in Stanford and suffers from muon induced neutron events.
Therefore a (not unproblematic) neutron subtraction is required to derive dark matter 
limits: 
using a Monte Carlo simulation of the neutron background, all candidate nuclear 
recoil events have been found to be compatible with the expected background. Within 
the year 2002 the experiment shall move to the Soudan deep underground lab.

The EDELWEISS collaboration is using 320\,g Ge detectors similar to those of CDMS, 
with NTD Ge phonon sensors and aluminium thin film charge collection electrodes, 
split on one crystal surface into a central electrode defining a fiducial volume 
and an outer ring electrode to reject events due to radioactive impurities near the 
detector circumference. The detectors are operated at $\sim$ 20\,mK in the Modane 
underground lab in the French-Italian alps. In data taking runs in the years 2000 
and 2002
(12.1\,kg$\cdot$days fiducial) no nuclear recoil events were observed in the relevant energy 
range in one of the detectors of each run, see fig.~\ref{fig:EDW}(a). 
A dark matter limit has been derived that 
excludes almost the entire DAMA allowed region,  
and the central value at 99.94\,\% CL \cite{EDW01}. 
At present this appears to be the most sensitive limit of all dark matter searches
for WIMP masses above 35\,GeV, 
moreover, due to the absence of any candidate events it is certainly the cleanest of 
all published limits. It is also the first experimental limit that excludes a small
part of the MSSM parameter space.
For an upgrade of the experiment to $\sim$~30\,kg a large 
volume reversed dilution refrigerator has been developed \cite{EDW02}.
The present limits of both the CDMS and EDELWEISS experiments are 
shown in fig.~\ref{fig:EDW}(b), together with the IGEX Ge diode limit and the 
allowed region of the DAMA WIMP candidate.

CRESST, CDMS and EDELWEISS expect, after upgrading their respective experiments to
$\sim 10 - 30$\,kg detector mass and improving background rejection efficiencies, very
similar sensitivities of about $10^{-8}\,$pb in the relevant WIMP mass range, thus
testing a significant part of MSSM models.

ROSEBUD, up to now using low threshold sapphire and Ge cryogenic detectors, are now
studying phonon-scintillation detectors and have obtained interesting results with $\sim 
50$\,g CaWO$_4$ and BGO crystals \cite{Cebrian01}. The recently terminated MIBETA 
experiment, conceived mainly for double-beta decay search and with $20 \times 340$\,g
TeO$_2$ detectors the hitherto biggest cryogenic experiment worldwide, is presently
being replaced by $56 \times 760$\,g TeO$_2$ (CUORICINO), later to be 
extended to 
$1000 \times 760$\,g TeO$_2$ (CUORE), however without background 
rejection \cite{CUORE01}. A Japanese collaboration is testing a $8 \times 21$\,g LiF 
crystal array, with the $^{19}$F nucleus sensitive to 
axial coupling, in the Kamioka underground lab \cite{Miuchi02}.

\subsection{More recent techniques}

The PICASSO \cite{PIC01} and SIMPLE \cite{SIMPLE} experiments are testing superheated Freon 
droplets, containing 
$^{19}$F and immersed in a gel matrix, as target material. A nuclear recoil inside a 
droplet above a 
certain threshold energy evaporates the droplet. The formation of a 
bubble can be registered as an acoustic pulse (``pop'') by a piezoelectric transducer. 
The threshold energy can be adjusted by varying the temperature. The 
advantage of this concept is, beside relative simplicity and cost efficiency, that 
the superheated droplets are insensitive to electron recoils with their much lower 
$dE/dx$ (energy dissipation per track length) such that the 
radioactive background problem essentially doesn't exist for these detectors, with 
the exception of alpha contaminants. Disadvantages are the small target masses 
employed up to now and the fact that the signals yield no information 
about the recoil energy.

The most promising attempt to reach directional sensitivity is the DRIFT-I detector 
in the Boulby mine, a low pressure CS$_2$ time projection chamber that allows imaging 
of recoil tracks \cite{Spooner02}. Beside a low sensitivity to electron recoils, the track 
length 
provides a means for discrimination. A problem is the low target density ($\sim 180\,$g in 
the 1\,m$^3$ DRIFT-I detector), i.e. the large volume required for a large-scale 
experiment. However, the convincing evidence that would arise from the observation of 
a modulation of recoil directions as expected from WIMP interactions seems attractive 
enough to pursue this development.

\section{Indirect WIMP searches}

Neutralinos can be gravitationally captured in celestial bodies. Since they are Majorana 
particles, pair 
annihilation can give rise to substancial fluxes of neutrinos, gammas or positrons 
\cite{Jung95}. Indirect WIMP searches are trying to detect these annihilation products.

In particular, indirect searches for 
neutralino annihilations inside the Sun, the Earth or the Galactic center are being 
undertaken by 
looking for neutrino fluxes resulting from the decay of gauge bosons and hadrons produced in
the annihilations. Muon neutrinos can be detected via the up-going muons
produced in charged-current interactions in the rock underneath
the big underground (Baksan, MACRO, SuperK) and 
underwater/ice (AMANDA, BAIKAL, ANTARES) experiments. Since both the neutrino-nucleon cross 
section and the muon range increase with the neutrino energy, the neutrino indirect 
detection of neutralinos improves with the parent neutralino mass.

\begin{figure}
\begin{center}
\begin{tabular}{ccc}
\mbox{\epsfig{file=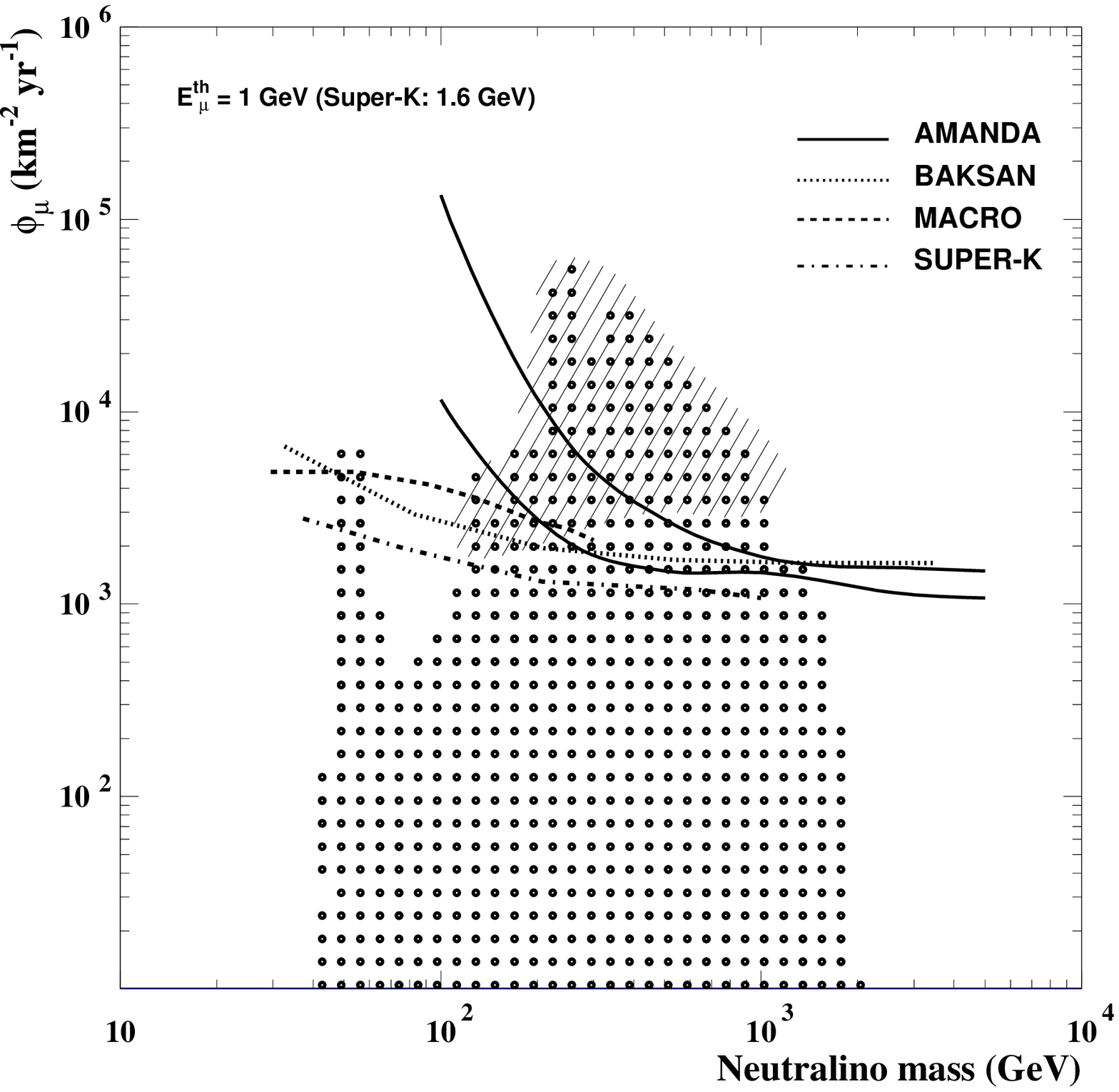,width=7.0cm}} &
\mbox{\hspace{0.3cm}} &
\mbox{\epsfig{file=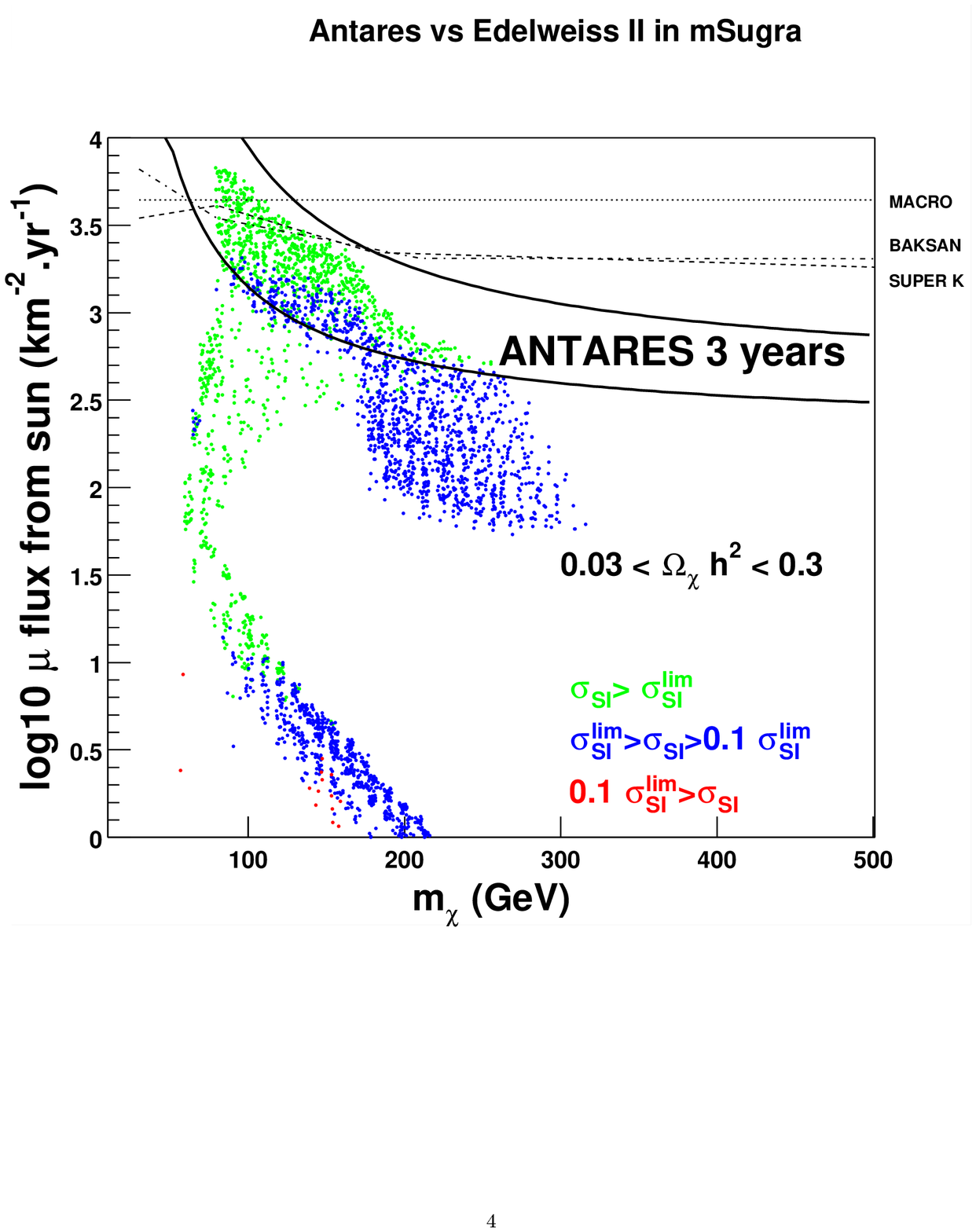,width=7.6cm}}
\end{tabular}
\caption{(a) Current exclusion limits of several indirect WIMP searches (lines) and model 
predictions (dots). The hatched region is disfavoured by direct searches. 
(b) Expected sensitivity of the ANTARES detector (solid lines) compared to current
limits of indirect searches. Different dot colours 
(tones of grey) show three years
sensitivity of the EDELWEISS II direct search, where $\sigma_{SI}$ denotes the WIMP-nucleon
cross section and $\sigma^{lim}_{SI}$ denotes 
the expected EDELWEISS II limit.}
\label{fig:indirect}
\end{center}
\end{figure}

Different frameworks have been considered to estimate such signals. A low energy
parameterization of the MSSM is usually used. Fig.~\ref{fig:indirect}(a) shows experimental 
limits of various indirect searches on the neutrino induced muon flux coming from the
center of the earch. The dots show model predictions from the MSSM framework estimated 
with the DarkSUSY package \cite{DARKSUSY}. 
Resonances in the cross section of 
neutralino capture in the center of the Earth could considerably enhance the signal, in 
particular at the mass of the Fe nucleus (56\,GeV), main constituent of the Earth's core.

A more theoretically motivated framework is provided by SuperGravity inspired models 
(CMSSM/ mSugra) based on the unification of the gauge couplings and masses at high scale and 
radiative electroweak symmetry breaking \cite{Cham82}.
Fig.~\ref{fig:indirect}(b) shows the expected 90\,\% CL sensitivity to the 
neutrino-induced muon flux coming 
from the Sun with the ANTARES 0.1 km$^2$ 10 string detector after three years of 
data taking as a function of the neutralino 
mass $m_{\chi}$ for a hard neutrino spectrum ($\chi\chi \rightarrow $ WW) \cite{Nezri02} 
(lower solid line), as well as for a soft spectrum (upper solid line), 
compared to current experimental limits. Also shown is the 
predicted signal from a wide range of mSugra
parameter space \cite{Bertin} calculated with the SUSPECT program \cite{SUSPECT}
for the SuSy particle 
spectrum and DarkSUSY for neutralino relic density and the neutrino/muon fluxes. Different 
point colours 
(tones of grey) show the sensitivity for three years of direct search by 
EDELWEISS II. 

Indirect detection of relic neutralino dark matter through annihilations into 
neutrinos, photons from the galactic center and from the galactic halo, and positrons has 
also been studied in 
ref. \cite{Feng,Ellis}.
Fluxes are computed for a set of benchmark CMSSM models. Gamma fluxes are compared to 
estimated sensitivities of the future experiments GLAST 
\cite{GLAST} and MAGIC \cite{MAGIC}. A cosmic ray positron excess observed by the HEAT
collaboration has been critically discussed in the light of a possible origin from 
WIMP annihilations in ref. \cite{Baltz}.

\section{Axions}

\begin{figure}
\begin{center}
\begin{tabular}{ccc}
\mbox{\epsfig{file=fig4a.eps,height=8cm}} &
\mbox{\hspace{1cm}} &
\mbox{\epsfig{file=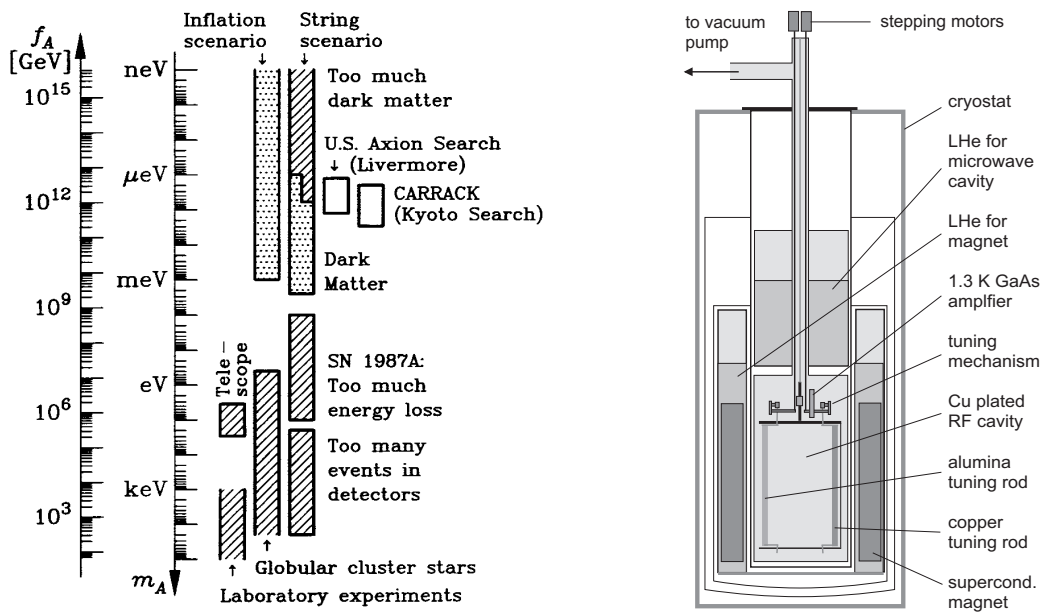,height=8cm}}
\end{tabular}
\caption{(a) Astrophysical and cosmological limits on the axion mass; the mass 
range sensitivities 
of the two principal axion search experiments are also shown. 
(b) Set-up of the U. S. Dark Matter Axion Search experiment.}
\label{fig:axion}
\end{center}
\end{figure}

The axion has been proposed by R. D. Peccei and H. R. Quinn as a solution to the 
strong CP problem in QCD: the CP violating electric dipole moment of the neutron
expected from the 
non-Abelian structure of QCD has been experimentally found to be suppressed to a 
vanishingly small value. This can be explained by the assumption of an additional 
global U(1) ``Peccei-Quinn'' symmetry. The explicit breaking of this symmetry is 
associated with a massive Goldstone boson, the axion. Dynamically produced in 
the early universe by relaxation of topological defects or by vacuum realignment - as 
opposed to the 
freeze-out from the primordial plasma of most other particles -, they constitute in 
spite of their small mass a CDM candidate. To date, the axion mass has been limited 
by astrophysical (too efficient cooling of SN\,1987\,A) and cosmological (too high 
axion DM density in the universe) constraints 
to $\sim 1\,\mu$eV $<$ m$_a$ $<$ 10\,meV, see fig.~\ref{fig:axion}(a) \cite{Raffelt}. 

Axions can couple to two photons via the Primakoff effect. This opens a way for direct 
axion detection. Sikivie proposed to provide virtual photons by means of a static 
magnetic field in a laboratory experiment to allow cosmic axions to convert to real 
photons that can be detected in a high-Q tunable microwave cavity. The open axion 
mass range can be covered by slowly scanning through the corresponding microwave 
frequency range. The coupling strength differs by roughly an order of magnitude for 
models with or without tree level lepton coupling (DFSZ or KSVZ axions, respectively).

The U. S. Dark Matter Axion Search \cite{Daw01} follows the proposal of Sikivie, 
using a single 
cavity of 50\,cm diameter and 1\,m length or a set of four cavities of 20\,cm diameter 
and 1\,m length fitting the same superconducting 7.6\,T magnet. Cavities and GaAs HFET 
electronics are cooled to 1.3\,K to reduce thermal noise. This configuration,
schematically represented in fig.~\ref{fig:axion}(b), is 
presently scanning part of the axion mass range at KSVZ sensitivity. In near future a 
newly developed SQUID based amplifier will be installed to further reduce the noise 
temperature, which should meet the requirements to search for KSVZ axions.

The CARRACK-II experiment \cite{Yam01} in Kyoto is also using a microwave cavity inside a 
superconducting magnet to convert axions to photons, however following a different 
strategy for photon detection. Microwaves are coupled to a separate, field-free 
detection cavity. Both cavities are cooled to 10\,mK. For photon detection a beam of 
two-step laser excited rubidium Rydberg atoms is passed through the detection cavity. 
Rydberg atoms whose excitation state has been enhanced by microwave photon absorption 
are selectively ionized by a precisely matched electric field, and the liberated 
electrons are detected in a channeltron electron multiplier. The single photon 
sensitivity should enable also this experiment to search for KSVZ axions.

\section*{Acknowledgments}
Valuable contributions by Dr. Sotiris Loucatos are acknowledged.


\section*{References}

\begin{thebibliography}{99}
\bibitem{Boerner93}G. B\"{o}rner, The Early Universe, Springer (1993)

\bibitem{Masi02}S. Masi {\em et al}, astro-ph/0201137 (2002)

\bibitem{Stompor01}R. Stompor {\em et al.}, Astrophys. J. 561 (2001) L7

\bibitem{Pryke02}C. Pryke {\em et al.}, Astrophys. J. 568 (2002) 46

\bibitem{Perl99}S. Perlmutter {\em et al.}, Astrophys. J. 517 (1999) 565

\bibitem{Efs01}G. Efstathiou {\em et al.}, astro-ph/0109152, subm. to MNRAS (2001)

\bibitem{Burles01}S. Burles {\em et al.}, Phys. Rev. D 63 (2001) 063512

\bibitem{Fukuda99}Y. Fukuda {\em et al.}, Phys. Rev. Lett. 82 (1999)2644

\bibitem{Ahmad01}Q. R. Ahmad {\em et al.}, Phys. Rev. Lett. 87 (2001) 071301

\bibitem{Lak02}I. Laktineh, hep-ex/0205088; these proceedings

\bibitem{Berg02}L. Bergstr\"{o}m, to appear in Proc. Neutrino2002, 
Munich, Germany, 25-30 May 2002 

\bibitem{Baudis98}L. Baudis {\em et al.}, Phys. Rev. D 59 (1998) 022001

\bibitem{Baudis01}L. Baudis {\em et al.}, Phys. Rev. D 63 (2001) 022001

\bibitem{Klapdor01}H. V. Klapdor-Kleingrothaus  {\em et al.}, hep-ph/0104028 (2001)

\bibitem{Morales02}A. Morales {\em et al.}, Phys. Lett. B 532 (2002) 8

\bibitem{Bernabei00}R. Bernabei {\em et al.}, Phys. Lett. B 480 (2000) 23

\bibitem{Gerbier99}G. Gerbier {\em et al.}, astro-ph/9902194 (1999)

\bibitem{Spooner98}N. Spooner, Pub. Boston, Particles, Strings and Cosmology
(1998) 130

\bibitem{Hart02}S. P. Hart, to appear in Proc. Dark Matter 2002, Marina del Rey, USA, 
20-22 Feb. 2002

\bibitem{Spooner02}N. Spooner, to appear in Proc. of School and Workshop on
Neutrino Particle Astrophysics, Les Houches, France, 21 Jan. - 1 Feb. 2002 
(http://leshouches.in2p3.fr)

\bibitem{Yoshida00}S. Yoshida {\em et al.}, Nucl. Phys. B (Proc. Suppl.) 87 (2000) 58

\bibitem{Hazama01}R. Hazama {\em et al.}, nucl-ex/0107001 (2001)

\bibitem{CRESST02}G. Angloher {\em et al.}, Astropart. Phys. (2002), in press

\bibitem{CRESST01}M. Altmann {\em et al.}, astro-ph/0106314 (2001)

\bibitem{CDMS}D. Abrams {\em et al.}, astro-ph/0203500, subm. to Phys. Rev. D (2002)

\bibitem{EDW01}A. Benoit {\em et al.}, astro-ph/0206271 , subm. to Phys. Lett. B (2002)

\bibitem{EDW02}A. Juillard {\em et al.}, to appear in Proc. DARK 2002, Cape Town, 
South Africa, 4 - 9 Feb. 2002

\bibitem{Cebrian01}S. Cebrian {\em et al.}, astro-ph/0112272, to appear in Proc. 
TAUP-01 (2002)

\bibitem{CUORE01}I. Irastorza {\em et al.}, hep-ph/0108146, subm. to Nucl. Instr. 
\& Meth. A (2002)

\bibitem{Miuchi02}K. Miuchi {\em et al.}, astro-ph/0204411, subm. to Astropart. 
Phys. (2002)

\bibitem{PIC01}L. Lessard {\em et al.}, Proc. DARK 2000, 
Ed. H. V. Klapdor-Kleingrothaus (2001) 604

\bibitem{SIMPLE}J. Collar {\em et al.}, Phys. Rev. Lett. 85 (2000) 3083;
J. Collar {\em et al.}, to appear in Proc. DARK 2002, Cape Town, 
South Africa, 4 - 9 Feb. 2002

\bibitem{Jung95}G. Jungman and M. Kamionkowski, Phys. Rev. D 51 (1995) 328

\bibitem{DARKSUSY}P. Gondolo {\em et al.}, DarkSUSY, 
http://www.physto.se/$\sim$edsjo/darksusy/ 

\bibitem{Cham82}A. H. Chamseddine {\em et al.}, Phys. Rev. Lett. 49 (1982) 970

\bibitem{Nezri02}E. Nezri, to appear in Proc. SUSY02, Hamburg, Germany, 17-23 June 2002

\bibitem{Bertin}V. Bertin {\em et al.}, hep-ph/0204135 (2002)

\bibitem{SUSPECT}A. Djouani {\em et al.}, SUSPECT, 
http://www.lpm.univ-montp2.fr:7082/$\sim$kneur/suspect.html


\bibitem{Feng}J. Feng {\em et al.}, astro-ph/0008115; Phys. Rev. D 63 (2001) 045024 

\bibitem{Ellis}J. Ellis {\em et al.}, astro-ph/0110225 (2001)

\bibitem{GLAST}GLAST collaboration, NIM A 466 (2001) 292

\bibitem{MAGIC}M. Martinez, MAGIC collaboration, Proc. 26th ICRC, Eds. D. Kieda 
{\em et al.}, vol. 5 (1999) 219

\bibitem{Baltz}E. A. Baltz {\em et al.}, astro-ph/0109318 (2001)

\bibitem{Raffelt}G. Raffelt, Stars as Laboratories for Fundamental Physics, 
University of Chicago Press (1996)

\bibitem{Daw01}E. Daw, Proc. IDM 2000, Eds. N. Spooner and 
V. Kudryavtsev, World Scientific (2001) 638

\bibitem{Yam01}K. Yamamoto {\em et al.}, Proc. DARK 2000, 
Ed. H. V. Klapdor-Kleingrothaus (2001) 638

\end{thebibliography}
\end{document}